\documentclass[journal]{IEEEtran}

\usepackage{amsmath}
\usepackage{amssymb}
\usepackage{amsfonts}
\usepackage{graphicx}
\usepackage{epsfig}
\usepackage{subfigure}
\usepackage{psfrag}
\usepackage{cite}
\usepackage{latexsym}
\usepackage{url}
\usepackage{color}
\usepackage{multirow}

\PassOptionsToPackage{bookmarks={false}}{hyperref}

\begin{document}
\title{Exploiting Physical-Layer Security for Multiuser Multicarrier Computation Offloading
\author{Jie~Xu and Jianping~Yao}
\thanks{The authors are with the School of Information Engineering, Guangdong University of Technology (e-mail: jiexu@gdut.edu.cn, yaojp@gdut.edu.cn). \emph{(Corresponding author: J. Yao.)}}
}
%\author{Jie Xu$^1$, Yong Zeng$^2$, and Rui Zhang$^2$\\
%$^1$School of Information Engineering, Guangdong University of Technology\\
%$^2$Department of Electrical and Computer Engineering, National University of Singapore\\
%E-mail:~jiexu@gdut.edu.cn,~\{elezeng,~elezhang\}@nus.edu.sg

%}

%\setlength{\textwidth}{7.1in} \setlength{\textheight}{9.7in}
%\setlength{\topmargin}{-0.8in} \setlength{\oddsidemargin}{-0.30in}

\maketitle

\begin{abstract}
This letter considers a mobile edge computing (MEC) system with one access point (AP) serving multiple users over a multicarrier channel, in the presence of a malicious eavesdropper. In this system, each user can execute the respective computation tasks by partitioning them into two parts, which are computed locally and offloaded to AP, respectively. We exploit the physical-layer security to secure the multiuser computation offloading from being overheard by the eavesdropper. Under this setup, we minimize the weighted sum-energy consumption for these users, subject to the newly imposed secrecy offloading rate constraints and the computation latency constraints, by jointly optimizing their computation and communication resource allocations. We propose an efficient algorithm to solve this problem.\vspace{-1em}
\end{abstract}

\newtheorem{definition}{\underline{Definition}}[section]
\newtheorem{fact}{Fact}
\newtheorem{assumption}{Assumption}
\newtheorem{theorem}{\underline{Theorem}}[section]
\newtheorem{lemma}{\underline{Lemma}}[section]
\newtheorem{corollary}{\underline{Corollary}}[section]
\newtheorem{proposition}{\underline{Proposition}}[section]
\newtheorem{example}{\underline{Example}}[section]
\newtheorem{remark}{\underline{Remark}}[section]
\newtheorem{algorithm}{\underline{Algorithm}}[section]
\newcommand{\mv}[1]{\mbox{\boldmath{$ #1 $}}}
\setlength\abovedisplayskip{0pt}
\setlength\belowdisplayskip{0pt}

\vspace{-1em}
\section{Introduction}
\vspace{-0.5em}
Mobile edge computing (MEC) has emerged as a promising technique to enhance the computation capacity and energy efficiency of wireless devices, for enabling various computation-intensive and latency-critical Internet-of-things (IoT) applications \cite{Mao2017}. By deploying MEC servers  at the network edge such as access points (APs), IoT devices can wirelessly offload the computation-heavy tasks to APs for efficient remote execution (see, e.g., \cite{BarbarossaSardellittiLorenzo2014,Chen2016,WangXuTWC,WangXuGCWorkshop,YouHuangChaeKim2017}). Despite the benefits, the wireless task offloading introduces new data security problems for wireless IoT devices. Due to the broadcast nature of wireless communications, the computation tasks offloaded from these devices are likely to be overheard by malicious attackers nearby, which may decode such information for launching security attacks. For the success of MEC, it is crucial to keep the confidentiality of the task offloading against eavesdropping attacks. %, to our best knowledge, this issue has not been addressed in the MEC literature yet.

Physical-layer security has emerged as a viable solution to ensure perfectly secured wireless communications against eavesdropping attacks, {provided that (partial) channel state information (CSI) of the eavesdroppers is available at the legitimate users} (see, e.g., \cite{Wyner1975,LiangPoorShamai2008,WangTao2011}). In physical-layer security, the key design objective is to maximize the so-called secrecy rate, i.e., the secure communication rate under the condition that the eavesdroppers cannot overhear any information.

\begin{figure}
\centering
 \epsfxsize=1\linewidth
    \includegraphics[width=6.4cm]{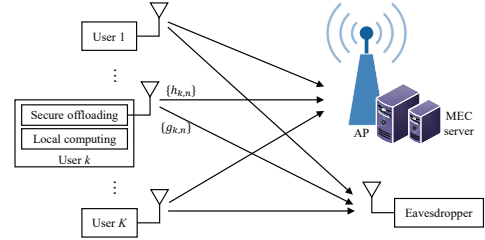}
\caption{The MEC system model with secure multiuser computation offloading over a multicarrier channel, in the presence of a malicious eavesdropper.} \label{fig0}\vspace{-2em}
\end{figure}

In this letter, we propose to employ the physical-layer security to secure the wireless computation offloading in MEC. We particularly focus on a multiuser multicarrier (e.g., orthogonal frequency-division multiple access (OFDMA)) system as shown in Fig. \ref{fig0}, in which a single AP (with an MEC server integrated) serves multiple users for their computation offloading, in the presence of a malicious eavesdropper. Each user can partition the computation tasks into two parts, which are computed locally and securely offloaded to AP, respectively. {Due to the employment of physical-layer security, new secure offloading constraints are imposed, i.e., the offloading rate at each user cannot exceed its secrecy rate to the AP, such that no information will be leaked to the eavesdropper.} {By taking into account such constraints and considering practical issues such as the imperfect CSI of the eavesdropper, how to jointly optimize the communication and computation resource allocations at multiple users for efficient MEC is a new problem that has not been investigated in the literature yet. Under this setup, we minimize the weighted sum-energy consumption for these users while ensuring their computation latency requirements, by jointly optimizing the users' local computing, as well as their transmit power and subcarrier allocations for secure offloading.} Although this problem is non-convex, we obtain its solution in a semi-closed form via the Lagrange duality method. Via numerical results, we validate the effectiveness of our proposed design over other benchmark schemes. We also show that as compared to the conventional setup without eavesdropper, the users consume more energy to secure the computation offloading from the eavesdropper's interception.

\vspace{-1em}
\section{System Model}
\vspace{-0.5em}
As shown in Fig. \ref{fig0}, we consider an MEC system with a single AP (with an MEC server integrated) and $K > 1$ users, in the presence of a malicious eavesdropper.{\footnote{Our results are extendible to the case with more than one eavesdropper, in which each user's achievable secrecy rate for offloading should be modified based on that in the so-called compound wire-tap channels with multiple eavesdroppers (see, e.g., \cite{Liang2009}).} Let $\mathcal K \triangleq \{1,\ldots,K\}$ denote the set of users. All nodes are equipped with a single antenna. We focus on a particular time block with duration $T$, during which each user $k\in\mathcal K$ needs to execute the computation tasks with $L_k > 0$ input bits. We consider the data partition task model for partial offloading, in which each task-input bit can be viewed as an independent sub-task. Therefore, user $k$ can partition the respective tasks into two portions with $l_k$ and $(L_k-l_k)$ input bits, which are locally computed at the user itself and securely offloaded to the AP over a multicarrier channel for remote execution, respectively. We consider a quasi-static subcarrier channel model, in which the wireless channels remain constant over each subcarrier within this block. Let $N$ denote the number of subcarriers in this system. For each subcarrier $n\in\mathcal N \triangleq \{1,\ldots,N\}$, let $h_{k,n}$ and $\tilde{g}_{k,n}$ denote the channel power gains from user $k$ to the AP and the eavesdropper, respectively. We assume that the AP perfectly knows the CSI of $h_{k,n}$'s and the computation information of all users, {but only partially knows that of $\tilde{g}_{k,n}$'s.{\footnote{When the eavesdropper is active, each user $k$ can monitor the eavesdropper's potential active transmission to estimate the corresponding $\tilde{g}_{k,n}$'s. When the eavesdropper is passive without transmitting any signal, each user $k$ can still detect the passive eavesdropping, by e.g. from the eavesdropper's local oscillator power leaked from its RF front end \cite{Mukherjee2012}. For both cases, the users can eventually obtain some information of $\tilde{g}_{k,n}$'s and then send such information back to the AP. In this case, the AP can partially know $\tilde{g}_{k,n}$'s with some errors.}} As commonly adopted in the physical-layer security literature \cite{WangSecure2013,Muhammad2017,Khandaker2018}, we consider the deterministic CSI uncertainty model for $\tilde{g}_{k,n}$'s, where $\tilde{g}_{k,n}=\overline{g}_{k,n}+\Delta g_{k,n}, k\in\mathcal K, n\in\mathcal N$. Here, $\overline{g}_{k,n}$ denotes the estimated CSI of $\tilde{g}_{k,n}$ at the AP and $\Delta g_{k,n}$ denotes the estimation error that is bounded by a possible value $\epsilon \ge 0$ (also know by the AP) as $|\Delta g_{k,n} | \leq \epsilon$.} %We also assume that the AP knows , in order to coordinate the joint local computing and secure offloading design.

%{\footnote{When the eavesdropper is active, each user $k$ can monitor the eavesdropper's potential active transmission to estimate the corresponding $g_{k,n}$'s and send such information back to the AP. For the purpose of initially investigating the physical-layer security in MEC, we make an ideal assumption that such CSI estimation is perfect at the users and the AP. Nevertheless, our results can be extended to the case when the CSI is not perfect, in which case the concepts of ergodic and outage secrecy rates (see, e.g., \cite{ZhouXiangYun2}) may be applicable. This, however, is beyond the scope of this work and left for future work. Furthermore, our design is also applicable in the case when the users and the AP can set a so-called secrecy guard zone (without any eavesdroppers in this zone) by detecting the existence of eavesdroppers in their vicinities \cite{ZhouXiangYun1}. In this case, the users and the AP can consider a robust physical-layer security design by setting $g_{k,n}$'s as the best possible channel power gains when the eavesdropper is outside this guard zone. }}

%Each user and the AP can share the CSI of $h_{k,n}$'s and the computation-related information via the reverser-link feedback. Depending on application scenarios, there have been various approaches for the AP to obtain the CSI of $g_{k,n}$'s. For example

As for the local computing of the $l_k$ input bits at each user $k\in\mathcal K$, let $C_k$ denote the number of CPU cycles required for computing {one task-input bit (or each independent sub-task). Accordingly, the total number of CPU cycles required for computing the $l_k$ bits is $C_k l_k$. Employing the dynamic voltage and frequency scaling technique \cite{Mao2017,BarbarossaSardellittiLorenzo2014}, user $k$ can control the CPU frequency $f_{k,m}$ for each cycle $m\in\{1,\ldots,C_k l_k\}$. In particular, in order to minimize the energy consumption for local computing at each user $k$, the CPU frequencies $f_{k,m}$'s should be identical over different cycles $m$'s \cite{WangXuTWC}. By using this fact and noting that the local execution time should be $T$ to meet the computation latency, we have the CPU frequencies at each user $k$ as $f_{k,m} = C_k l_k/T, \forall m\in\{1,\ldots,C_k l_k\}$. Therefore, the user $k$'s energy consumption for local computing is given by
%\begin{align}
$E_k^{\text{loc}} = \sum_{m=1}^{C_kl_k} \zeta_k f_{k,m}^2 = \zeta_k C_k^3 l_k^3/T^2$,
%\end{align}
where $\zeta_k$ denotes the effective capacitance coefficient that depends on the chip architecture at user $k$ \cite{Mao2017}. Furthermore, let $f_k^{\max}$ denote the maximum CPU frequency at each user $k$; we have $f_{k,m} \le f_k^{\max}, \forall k,m$. Accordingly, it must hold that $l_k \le l_k^{\max} \triangleq f_k^{\max}T/C_k, \forall k\in\mathcal K$.

Next, we consider the secure offloading of the $(L_k - l_k)$ task input bits for each user $k\in\mathcal K$.\footnote{Note that the MEC server and the AP generally have large computation capability and transmission power, respectively. Therefore, we ignore the time required for remote computation at the MEC server and  computation results downloading from the AP to the users (see, e.g., \cite{WangXuTWC,WangXuGCWorkshop,YouHuangChaeKim2017}).} Let $\{\theta_{k,n}\} $ denote the indicators for subcarrier allocation with $\theta_{k,n}\in \{0,1\}$, where $\theta_{k,n} = 1$ or $\theta_{k,n} = 0$ mean that the sub-carrier $n\in\mathcal N$ is or is not allocated to user $k$, respectively. Let $p_{k,n} \ge 0$ denote the transmit power at user $k$ for secure task offloading, and $B$ the bandwidth of each subcarrier. {Under the CSI uncertainty model, the worst-case achievable secrecy rate (in bits/sec) at user $k$ for offloading is given as %\cite{Wyner1975}
%\begin{footnotesize}
\begin{align*}
&R_k(\mv \theta_k,\mv p_k) = \min_{|\Delta g_{k,n}| \le \epsilon}
B \sum_{n=1}^N \theta_{k,n}\big(\log_2\big(1 + h_{k,n} p_{k,n}\big) \\
&- \log_2\big(1 + {\tilde{g}_{k,n} p_{k,n}}\big)\big)^+\end{align*}
\begin{align*}
&=
B \sum_{n=1}^N \theta_{k,n}\big(\log_2\big(1 + h_{k,n} p_{k,n}\big)
- \log_2\big(1 + {g_{k,n} p_{k,n}}\big)\big)^+,
\end{align*}where $g_{k,n} = \overline{g}_{k,n} + \epsilon$ denotes the best possible channel power gain of the eavesdropper known by the AP.} Here, the receiver noise powers at the AP and the eavesdropper are normalized to be unity, $(x)^+ \triangleq \max(x,0)$, $\mv \theta_k \triangleq [\theta_{k,1},\ldots,\theta_{k,N}]^\dagger$, and $\mv p_k \triangleq [p_{k,1},\ldots,p_{k,N}]^\dagger$, with the superscript $\dagger$ denoting the transpose. The user $k$'s transmission energy consumption for secure offloading is given as
%\begin{align}
$E_k^{\text{off}} = \sum_{n=1}^N \theta_{k,n} p_{k,n}T$.
%\end{align}
%\end{footnotesize}

Under this setup, our objective is to minimize the weighted sum-energy consumption at the $K$ users (i.e., $\sum_{k=1}^K \alpha_k(E_k^{\text{loc}} + E_k^{\text{off}})$) while ensuring the successful computation task execution within this block. Here, $\alpha_k > 0$ denotes the energy weight for each user $k\in\mathcal K$, where a larger value of $\alpha_k$ indicates a higher priority for user $k$ in energy minimization. The decision variables include the task partition $\mv l \triangleq [l_{1},\ldots,l_{K}]^\dagger$, as well as the subcarrier allocation $\mv{\Theta} \triangleq [\mv \theta_1,\ldots,\mv \theta_{K}]$ and the power allocation $\mv{P} \triangleq [\mv p_1,\ldots,\mv p_{K}]$ for secure task offloading. Mathematically, this problem is formulated as
%\begin{footnotesize}
\begin{align}
\mathrm{(P1)}:&\min_{\mv l, \mv{\Theta},\mv P} \sum_{k=1}^K \alpha_k \bigg(\zeta_k C_k^3 l_k^3/T^2 + \sum_{n=1}^N \theta_{k,n} p_{k,n}T\bigg) \nonumber\\
\mathrm{s.t.}~& TR_k(\mv \theta_k,\mv p_k) \ge L_k - l_k, \forall k\in\mathcal K \label{eqn:rate:con}\\
&0\le l_k \le l_k^{\max}, p_{k,n} \ge 0, \forall k\in\mathcal K, n\in\mathcal N\label{eqn:power:con}\\
&\sum_{k=1}^K \theta_{k,n} = 1, \theta_{k,n} \in \{0,1\},\forall n\in\mathcal N,\label{eqn:subcarrier:con}
\end{align}
%\end{footnotesize}
{Notice that in \eqref{eqn:rate:con}, the worst-case secrecy rate for each user $k$ must be no smaller than the offloading rate, such that the offloading is secured under any possible eavesdropper channels.} Furthermore, the constraints in \eqref{eqn:subcarrier:con} ensure that each subcarrier is only allocated to one user. However, due to the binary variables in $\mv{\Theta}$, problem (P1) is a non-convex optimization problem that is generally difficult to solve.

{Before proceeding, it is worth noting that in the special case without the eavesdropper (or equivalently $g_{k,n} = 0, \forall k\in\mathcal K,n\in\mathcal N$), problem (P1) corresponds to the energy-efficient multiuser computation offloading problem over multicarrier systems in \cite{YouHuangChaeKim2017}. In the other special case with only offloading (or equivalently $l_{k} = 0, \forall k\in\mathcal K$), problem (P1) corresponds to a secrecy communication problem over a multicarrier channel (see, e.g., \cite{WangTao2011}). Therefore, problem (P1) unifies the conventional computation offloading design in MEC and the energy efficient communication with physical-layer security.}
\vspace{-1em}
\section{Proposed Solution to Problem (P1)}
\vspace{-0.5em}
Though non-convex, it can be shown that problem (P1) satisfies the time-sharing condition in \cite{YuLui2006}, as the number of subcarriers $N$ becomes infinite. In this case, zero duality gap or strong duality holds between (P1) and its Lagrange dual problem. In this section, we solve problem (P1) by using the Lagrange dual method, by considering the zero duality gap.\footnote{In our simulations in Section \ref{sec:IV} with $N=64$ subcarriers, the duality gap of (P1) is actually negligibly small and thus can be ignored. Moreover, the duality gap reduces as $N$ increases, and approaches zero for $N \to \infty$\cite{YuLui2006}.}

Let $\lambda_k \ge 0, k\in\mathcal K$, denote the dual variable associated with the $k$-th constraint in \eqref{eqn:rate:con}. The Lagrangian of (P1) is given as\vspace{-0.9em}

\begin{footnotesize}
\begin{align}
&\mathcal{L}(\mv l, \mv{\Theta},\mv P,\mv \lambda)=  \sum_{k=1}^K \alpha_k \big(\zeta_k C_k^3 l_k^3/T^2 + \sum_{n=1}^N \theta_{k,n}p_{k,n}T\big)\\
\nonumber&- \sum_{k=1}^K \lambda_k\bigg(T B\sum_{n=1}^N \theta_{k,n}\left(\log_2 \frac{1 + h_{k,n} p_{k,n}}{1 + g_{k,n} p_{k,n}}\right)^+  - (L_k - l_k)
\bigg).
\end{align}
\end{footnotesize}
The dual function is given by
\begin{align}
f(\mv \lambda) = \min_{\mv l, \mv{\Theta},\mv P}& \mathcal{L}(\mv l, \mv{\Theta},\mv P,\mv \lambda),~~
\mathrm{s.t.}~\eqref{eqn:power:con}~{\text{and}}~\eqref{eqn:subcarrier:con}.\label{eqn:dual:function}
\end{align}
The dual problem is
\begin{align}
\mathrm{(D1)}:\max_{\mv \lambda}~&f(\mv \lambda),~\mathrm{s.t.}~\lambda_k \ge 0, \forall k\in\mathcal K.\label{eqn:dual:problem}
\end{align}
In the following, we solve problem (P1) by first solving problem \eqref{eqn:dual:function} to obtain $f(\mv \lambda)$ under any given $\mv \lambda$ satisfying \eqref{eqn:dual:problem}, and then solving (D1) via updating $\mv \lambda$ to maximize $f(\mv \lambda)$.

First, consider problem \eqref{eqn:dual:function} under any given $\mv \lambda$ satisfying \eqref{eqn:dual:problem}. In this case, problem \eqref{eqn:dual:function} can be decomposed into the following subproblems by dropping the irrelevant constant $\sum_{k=1}^K\lambda_k L_k$.
\begin{align}
&~~~~\min_{0\le l_k \le l_k^{\max}}~\alpha_k \zeta_k C_k^3l_k^3/T^2 - \lambda_k l_k, \label{eqn:sub:problem:1}\\
&\min_{\{\theta_{k,n}, p_{k,n}\}_{k=1}^K} ~\sum_{k=1}^K\alpha_k  \theta_{k,n}p_{k,n}T - \sum_{k=1}^K \lambda_k
\theta_{k,n}TB\nonumber\\
&~~~~\times\big(\log_2\big(1 + {h_{k,n} p_{k,n}}\big)- \log_2\big(1 + {g_{k,n} p_{k,n}}\big)\big)^+\nonumber\\
&~~\mathrm{s.t.}~p_{k,n} \ge 0, \theta_{k,n} \in \{0,1\}, \forall k\in\mathcal K,~ \sum_{k=1}^K \theta_{k,n} = 1,\label{eqn:sub:problem:2}
\end{align}
where each subproblem \eqref{eqn:sub:problem:1} corresponds to one user $k$, and each subproblem \eqref{eqn:sub:problem:2} corresponds to one subcarrier $n$.

For the $k$th subproblem in \eqref{eqn:sub:problem:1}, by checking the first-order derivative, the optimal solution is given by
\begin{align}\label{eqn:l:k}
{l_k^{(\mv \lambda)} = \min\left(\sqrt{\frac{\lambda_k T^2}{3\alpha_k \zeta_kC_k^3}},l_k^{\max}\right).}
\end{align}

For the $n$th subproblem in \eqref{eqn:sub:problem:2}, it is evident that only one user can be active due to the constraint $\sum_{k=1}^K \theta_{k,n} = 1$, i.e., there exists exactly one user $k$ such that $\theta_{k,n} = 1$ and $\theta_{\hat k,n} = 0, \forall \hat k \neq k$. As a result, we can optimally solve this problem by solving for $\{p_{k,n}\}_{k=1}^K$ under each possible $\{\theta_{k,n}\}_{k=1}^K$, and then comparing the resultant objective values to find the optimal $\{\theta_{k,n}\}_{k=1}^K$. When user $k$ is active (i.e., $\theta_{k,n} = 1$ and $\theta_{\hat k,n} = 0, \forall \hat k \neq k$), we define the objective function of problem \eqref{eqn:sub:problem:2} as
\begin{align}
\nonumber\psi_{k,n}(p_{k,n})\triangleq \alpha_k p_{k,n}T - \lambda_k TB \left(\log_2\frac{1 + h_{k,n} p_{k,n}}{1 + g_{k,n} p_{k,n}}\right)^+.
\end{align}
Then we have the following lemma to solve problem \eqref{eqn:sub:problem:2}.
\begin{lemma}\label{lemma:1}
For the $n$th subproblem in \eqref{eqn:sub:problem:2} under given $\mv\lambda$, the optimal power allocation solution is given as
\begin{align}\label{eqn:p_kn:lambda}
p_{k,n}^{(\mv\lambda)}= &\left\{
\begin{array}{ll}
0, & {\text{if}}~h_{k,n} \le g_{k,n} \\
\left(\frac{\lambda_kB}{\ln 2 \alpha_k} - \frac{1}{h_{k,n}}\right)^+& {\text{if}}~g_{k,n}=0\\
\bigg(\frac{\sqrt{\Delta^{(\mv\lambda)}_{k,n}}-(h_{k,n} + g_{k,n})}{2h_{k,n}g_{k,n}}
\bigg)^+ & {\text{otherwise}},
\end{array}
\right.
\end{align}
$\forall k\in\mathcal K$, where \begin{align}%\label{eqn:Delta}
\nonumber\Delta_{k,n}^{(\mv\lambda)} = (h_{k,n} - g_{k,n})^2 + \frac{4\lambda_kBh_{k,n}g_{k,n}}{\ln 2\alpha_k} (h_{k,n} - g_{k,n}).
\end{align}
In this case, the index of the active user is
\begin{align}\label{eqn:k:lambda:n}
k^{(\mv\lambda)}_n = \arg \min_{k\in\mathcal K} \psi_{k,n}(p_{k,n}^{(\mv\lambda)}),
\end{align}
and accordingly, the subcarrier allocation is given as
\begin{align}\label{eqn:theta:lambda:n}
\theta_{k,n}^{(\mv\lambda)} = & \left\{
\begin{array}{ll}
1, & {\text{if}}~k= k^{(\mv\lambda)}_n \\
0,& {\text{otherwise}}
\end{array}
\right.,\forall k\in\mathcal K.
\end{align}
\end{lemma}

\begin{IEEEproof}
%See Appendix \ref{app:A}.
Suppose that user $k$ is active with $\theta_{k,n} = 1$ and $\theta_{\hat k,n} = 0, \forall \hat k \neq k$, problem \eqref{eqn:sub:problem:2} is reexpressed as
$\min_{p_{k,n} \ge 0} \psi_{k,n}(p_{k,n})$.
%\end{align}
When $h_{k,n} \le g_{k,n}$, we have $\log_2(1 + h_{k,n} p_{k,n}) - \log_2(1 + g_{k,n} p_{k,n})\le 0$ under any $p_{k,n}\ge 0$, and therefore, it follows that $p_{k,n}^{(\mv\lambda)} = 0$ in this case. When $h_{k,n} > g_{k,n}$, this problem is indeed convex. By checking the first-order derivative of $\psi_{k,n}(p_{k,n})$ in this case, we have $p_{k,n}^{(\mv\lambda)}$ in \eqref{eqn:p_kn:lambda}. As a result, the optimal objective value of problem \eqref{eqn:sub:problem:2} in the case with the user $k$ being active is given as $\psi_{k,n}(p_{k,n}^{(\mv\lambda)})$. By comparing $\psi_{k,n}(p_{k,n}^{(\mv\lambda)})$'s under different $k$'s, the optimal $k^{(\mv\lambda)}_n$ and $\theta_{k,n}^{(\mv\lambda)}$ can be obtained in \eqref{eqn:k:lambda:n} and \eqref{eqn:theta:lambda:n}, respectively. %Therefore, this lemma is verified.
\end{IEEEproof}
%
%\begin{align}
%\theta_{k,n}^{(\mv\lambda)} = & \left\{
%\begin{array}{ll}
%1, & {\text{if}}~k\neq k^*_n \\
%0,& {\text{otherwise}}.
%\end{array}
%\right.
%\\
%p_{k,n}^*= &\left\{
%\begin{array}{ll}
%0, & {\text{if}}~h_{k,n} \le g_{k,n} \\
%\left(\frac{\lambda_kB}{\ln 2 \alpha_kT} - \frac{1}{h_{k,n}}\right)^+& {\text{if}}~g_{k,n}=0\\
%\left(\frac{-(h_{k,n} + g_{k,n}) \pm \sqrt{\Delta_{k,n}})}{2h_{k,n}g_{k,n}}
%\right)^+ & {\text{otherwise}},
%\end{array}
%\right.
%\end{align}
%where the index of the active user is given as
%\begin{align}
%\min_{p_{k,n} \ge 0} &~\psi_{k,n}(p_{k,n}),
%\end{align}

By combining $l_k^{(\mv \lambda)}$'s in \eqref{eqn:l:k} as well as $\theta_{k,n}^{(\mv\lambda)}$'s and $p_{k,n}^{(\mv\lambda)}$'s in Lemma \ref{lemma:1}, the dual function $f(\mv \lambda)$ in \eqref{eqn:dual:function} is obtained.

Next, it remains to solve problem (D1). As the dual problem (D1) is always convex but generally non-differentiable, we can use subgradient-based methods such as the ellipsoid method to solve (D1) optimally, by using the fact that the subgradient of $f(\mv \lambda)$ with respect to $\lambda_k$ is $( L_k - l_k^{(\mv\lambda)}) - TR_k(\mv \theta_k^{(\mv\lambda)},\mv p_k^{(\mv\lambda)})$. We denote $\mv\lambda^{*} = [\lambda_1^{*},\ldots,\lambda_K^{*}]^\dagger$ as the optimal dual solution to (D1).

Finally, based on the optimal $\mv\lambda^{*}$ to (D1), we have the following proposition to solve (P1).

\begin{proposition}\label{proposition:1}
The solution to problem (P1) is given as $l_k^* = l_k^{(\mv\lambda^*)}, p_{k,n}^* = p_{k,n}^{(\mv\lambda^*)}$, and $\theta_{k,n}^* = \theta_{k,n}^{(\mv\lambda^*)}, \forall k\in\mathcal K, n\in\mathcal N$, where $l_k^{(\mv \lambda)}$'s, $p_{k,n}^{(\mv\lambda)}$'s, and $\theta_{k,n}^{(\mv\lambda)}$'s are given in \eqref{eqn:l:k}, \eqref{eqn:p_kn:lambda}, and  \eqref{eqn:theta:lambda:n}, respectively.
%
%
%\begin{align}
%l_k^* = & \sqrt{\frac{\lambda_k^* T^2}{2\alpha_k \zeta_k}}\\
%\theta_{k,n}^* = & \left\{
%\begin{array}{ll}
%1, & {\text{if}}~k\neq k^*_n \\
%0,& {\text{otherwise}}.
%\end{array}
%\right.
%\\
%p_{k,n}^*= &\left\{
%\begin{array}{ll}
%0, & {\text{if}}~h_{k,n} \le g_{k,n} \\
%\left(\frac{\lambda_kB}{\ln 2 \alpha_kT} - \frac{1}{h_{k,n}}\right)^+& {\text{if}}~g_{k,n}=0\\
%\left(\frac{-(h_{k,n} + g_{k,n}) \pm \sqrt{\Delta_{k,n}})}{2h_{k,n}g_{k,n}}
%\right)^+ & {\text{otherwise}},
%\end{array}
%\right.
%\end{align}
%Here, we have \begin{align}
%k^*_n = \arg \min_{k\in\mathcal K} \psi_{k,n}(p_{k,n}^*).
%\end{align}
%$\mv \lambda^* = [\lambda_1^*,\ldots,\lambda_K^*]$ corresponds to the optimal dual variables that can be obtained based on the ellipsoid method.
\end{proposition}
%\begin{IEEEproof}
%See Appendix.
%\end{IEEEproof}

\begin{remark}
%It is worth noting that the semi-closed-form solution to (P1) in Proposition \ref{proposition:1} provides insights on the joint local computing and secure offloading design at these users. %First, the dual variable $\lambda^*$ plays the role in controlling the energy consumption tradeoff between the local computing and secure offloading.
Proposition \ref{proposition:1} generalizes the resource allocation for the computation offloading in MEC and that for the physical-layer security (see, e.g., \cite{WangTao2011}) over multicarrier systems. First, when $g_{k,n} = 0,\forall k\in\mathcal K,n\in\mathcal N$, a water-filling-like power allocation is observed in \eqref{eqn:p_kn:lambda} for each user. This corresponds to the energy efficient multiuser computation offloading design without eavesdropper (see, e.g., \cite{YouHuangChaeKim2017}). Next, it is observed that the optimal power and subcarrier allocations in \eqref{eqn:p_kn:lambda} and \eqref{eqn:theta:lambda:n} have similar structures as those for the secrecy communication over a multicarrier channel in \cite{WangTao2011}, while the difference lies in the determination of the dual variable $\lambda^*$, which controls the energy consumption tradeoff between the local computing and secure offloading in our consideration.

%, in which the allocated power $p_{k,n}^*$ increases monotonically with respect to the channel power gain $h_{k,n}$ of the communication channel but decreases monotonically with respect to $g_{k,n}$ of the eavesdropping channel. %Therefore, the solution in Proposition \ref{proposition:1} indeed unifies the optimal resource allocation designs for conventional secrecy communication without computation, as well as conventional computation offloading design without security consideration.
\end{remark}
\vspace{-1em}
\section{Numerical Results}\label{sec:IV}
\vspace{-0.5em}
In this section, we present numerical results to validate the performance of our proposed design as compared to two benchmark schemes, as well as the conventional design without eavesdropper that servers as a performance upper bound (or energy lower bound).
\subsubsection{Secure full offloading}
All the $K$ users choose to offload all the task input bits to the AP. In this case, the weighted sum-energy minimization corresponds to solving problem (P1) by setting $l_k = 0, \forall k\in\mathcal K$.
\subsubsection{Local computing}
All the $K$ users locally compute all the computation tasks, i.e., $l_k = L_k, \forall k\in\mathcal K$. The weighted sum-energy consumption by the $K$ users is $\sum_{k=1}^K \alpha_k \zeta_k C_k^3L_k^3/T^2$.
\subsubsection{Conventional design without eavesdropper}
The weighted sum-energy minimization corresponds to solving problem (P1) by setting $g_{k,n} = 0, \forall k\in\mathcal K, n\in\mathcal N$.

In the simulation, we consider a multicarrier system with $N=64$ subcarriers and $K=4$ users. We consider the Rayleigh fading channel model for $h_{k,n}$'s and $g_{k,n}$'s, and assume that the average channel power gains follow the pathloss model $\beta_0 (d/d_0)^{-\xi}$, where $d$ denotes the distance between the respective nodes, $\beta_0 = -30$ dB corresponds to the pathloss at a reference distance of $d_0 = 1$ meter (m), and $\xi = 3.7$ corresponds to the pathloss exponent. We set $\zeta_k = 10^{-28}$ Joule (J)/cycle, $C_k = 10^3$ cycles/bit, $B = 0.3125$ MHz, the noise power spectrum density to be $-105$ dBm/Hz, and {$\epsilon$ to be 10\% of the corresponding pathloss}. We also set $\alpha_k = 1/K, \forall k\in\mathcal K$, and thus we consider the average energy consumption at the $K$ users as the performance metric. We also consider $L_k = L, \forall k\in\mathcal K$, and set the distances from the $K$ users to the AP to be identical as $20$ meters.

\begin{figure}[!t]
\centering
 \epsfxsize=1\linewidth
    \includegraphics[width=5.5cm]{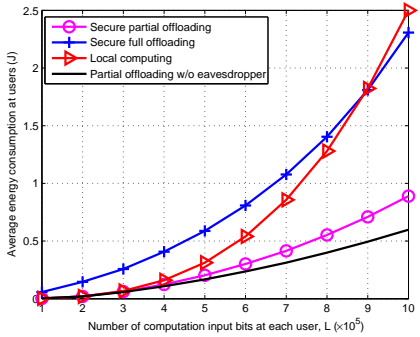}
\caption{{Average energy consumption at the users versus the number of computation input bits $L$ at each user.}} \label{fig1}\vspace{-2em}
\end{figure}

Fig. \ref{fig1} shows the average energy consumption of the $K$ users versus the number of computation input bits $L$ at each user, in which the distances from the $K$ users to the eavesdropper are all $20$ meters. It is observed that when $L$ is small (e.g., $L \le 3\times 10^5$ bits), the proposed design, the local computing, and the conventional design without eavesdropper achieve similar energy consumption performance, and outperform the secure full offloading. This is because in this case, the local computing is sufficient to handle the computation tasks. By contrast, when $L$ becomes large (e.g., $L \ge 4\times 10^5$ bits), the proposed design is observed to outperform the secure full offloading and local computing. This shows the importance of joint optimization of local computing and secure offloading. In this case, the proposed design is also observed to consume more energy than the conventional design without eavesdropper, for the purpose of anti-eavesdropping.

Fig. \ref{fig2} shows the average energy consumption of the $K$ users versus the identical distance from the users to the eavesdropper, in which we set $L = 7\times10^5$ bits. It is observed that as the distance increases, the energy consumption for secure offloading decreases, as the wireless channels to the eavesdropper become weaker. More specifically, the proposed design is observed to have a similar performance as the conventional design without eavesdropper, when the distance is larger than $30$ m.

%When the distance from users to the eavesdropper is equal to that to the AP (i.e., 200 m), it is observed that the local computing outperforms the secure full offloading, while the proposed design performs considerably inferior to the conventional design without eavesdroppers. This is due to the fact that in this case, a large amount of energy is consumed by these users for combating the eavesdropping. When the distance from users to the eavesdropper becomes larger, the energy consumption by both the secure full offloading and the proposed design are observed to reduce significantly.

\begin{figure}[!t]
\centering
 \epsfxsize=1\linewidth
    \includegraphics[width=5.5cm]{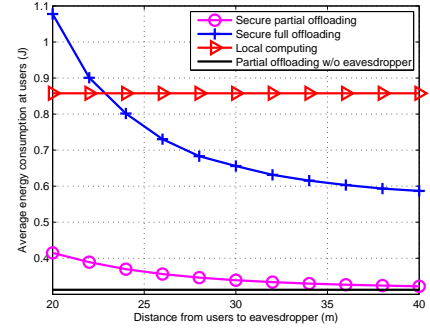}
\caption{{Average energy consumption at the users versus the distance from the users to the eavesdropper.}} \label{fig2}\vspace{-2em}
\end{figure}

\vspace{-1em}
\section{Conclusion}
\vspace{-0.5em}
This letter proposed to use physical layer security to ensure the computation task offloading in MEC systems. By focusing on a multiuser multicarrier system, we studied a latency-constrained weighted sum-energy minimization problem via jointly optimizing the local computing and secure offloading. How to extend the secure computation offloading to other MEC setups with, e.g., multiple APs and multiple antennas is interesting future directions worth investigating.

\vspace{-0.5em}
\footnotesize
\bibliographystyle{IEEEtran}
\vspace{-0.5em}
\bibliography{myreference}

\end{document}